\documentclass[10pt]{article}
\usepackage{amssymb}

\newcommand{\ket}[1]{\ensuremath{\left|  #1 \right\rangle}}

\newcommand{\op}[1]{\ensuremath{\widehat{\textsf{\ensuremath{#1}}}}}
\newcommand{\be}{\begin{equation}}
\newcommand{\ee}{\end{equation}}
\usepackage[colorlinks=true, pdfstartview=FitV, linkcolor=black, citecolor=black, urlcolor=blue]{hyperref}

\long\def\symbolfootnote[#1]#2{\begingroup%
\def\thefootnote{\fnsymbol{footnote}}\footnote[#1]{#2}\endgroup}

\title{Probability in the Everett World: Comments\\on Wallace and  Greaves\footnote{This note is prompted by several recent papers by David Wallace and Hilary Greaves. I hope it will be clear to readers who know those papers that while I focus (of course) on aspects of Wallace's and Greaves's arguments that I find problematic, I'm very much indebted to their discussions, which greatly clarify the issue of probability in the Everett world. }}

\author{Huw Price\footnote%
{\href{http://www.usyd.edu.au/time/}{Centre for Time}, Department of Philosophy, Main Quad A14, University of Sydney, NSW 2006, Australia. E-mail: \href{mailto:huw@mail.usyd.edu.au}{huw@mail.usyd.edu.au}.}}

\date{\today
}
\begin{document}
\maketitle
\begin{abstract}
\noindent It is often objected that the Everett interpretation of QM cannot make sense of quantum probabilities, in one or both of two ways: either it can't make sense of probability at all, or it can't explain why probability should be governed by the Born rule. David Deutsch has attempted to meet these objections. He argues not only that rational decision under uncertainty makes sense in the Everett interpretation, but also that under reasonable assumptions, the credences of a rational agent in an Everett world should be constrained by the Born rule.

David Wallace  has developed and defended Deutsch's proposal, and greatly clarified its conceptual basis.  In particular, he has stressed its reliance on the distinguishing symmetry of the Everett view, viz., that all possible outcomes of a quantum measurement  are treated as equally real. The argument thus tries to make a virtue of what has usually been seen as the main obstacle to making sense of probability in the Everett world. 

In this note I outline some objections to the Deutsch-Wallace argument, and  to related proposals by Hilary Greaves about the epistemology of Everettian QM. (In the latter case, my arguments include an appeal to an Everettian analogue of the Sleeping Beauty problem.) The common thread to these objections is that the symmetry in question remains a very significant obstacle to making sense of probability in the Everett interpretation.  \end{abstract}

\section{Preamble}

It is often objected that the Everett view of quantum theory cannot make adequate sense of quantum probabilities, in one or both of two senses: either it cannot make sense of probability at all, or cannot explain why probability should be governed by the Born rule. David Deutsch (1999) has attempted to meet these objections. He argues not only that rational decision under uncertainty (of the kind traditionally associated with probability) makes sense in the Everett interpretation; but also that under reasonable assumptions, the credences of a rational agent in an Everett world should be constrained by the Born rule.

In two recent papers (2003, 2005b), David Wallace  has developed and defended Deutsch's proposal, and greatly clarified its conceptual basis.  In this note, however, I want to outline  some concerns about the latest form of the Deutsch-Wallace (DW) argument, and related matters. In particular,  I want to argue that the argument is circular, at a crucial point. Similar objections have been raised to various other proposals to derive meaningful probabilities in the Everett context. The present objection  is related to reservations about the DW argument expressed by Barnum \emph{et al} (2000) and by Lewis (2003, 2005), though I think it develops these reservations in a new and more forceful way.

Most of my concerns turn  on a feature of the Everett view that Wallace takes to be crucial to the DW argument.  As he notes, the argument is based on symmetry considerations. He argues that it is stronger than the usual kind of symmetry-based attempts to  derive probabilities, because the symmetry isn't broken by a single \emph{actual} outcome (which, if one knew it, would trump the symmetry-based probabilities as a guide to action). Here is Wallace's own summary of the point from  another paper: 

\begin{quotation}
\noindent I will not attempt to summarise these decision-theoretic proofs here, since 
the details are somewhat involved, but the underlying principle is essentially 
that of symmetry: if there is a physical symmetry between two possible outcomes there can be no reason to prefer one to another. Such arguments have 
frequently been advanced in non-quantum contexts but ultimately fall foul of 
the problem that the symmetry is broken by one outcome rather than another 
actually happening (leading to a requirement for probability to be introduced 
explicitly at the level either of the initial conditions or of the dynamics to select 
which one happens). They find their natural home -- and succeed! -- in Everettian quantum mechanics, where all outcomes occur and there is no breaking 
of the symmetry. (Wallace, 2005a, \S3.6)
\end{quotation}
As I'll explain below, I have two main concerns related to this feature of the model. Together, these concerns  suggest that this supposed advantage is at best two-edged: the source of the symmetry is also the source of the deepest reasons for scepticism about the fate of probability in the Everett interpretation.

I think that the concerns I am going to identify also create difficulties for Hilary Greaves's proposed interpretation of Everettian probabilities (Greaves 2004). However, I shall be relying on Greaves's discussion and conclusions in other ways. In particular, I follow Greaves in rejecting what Wallace refers to as the  \emph{Subjective-uncertainty (SU) viewpoint.} Wallace characterises this viewpoint as follows:  
\begin{quote}
Given that what
it is to have a future self is to be appropriately related to a certain
future person, and that in normal circumstances I expect to become my
future self, so also in Everettian splittings I should expect to become \emph{one} 
of my future selves. If there is more than one of them I should be uncertain as to which I will
become; furthermore, this subjective uncertainty is compatible with my \emph{total
knowledge} of the wavefunction and its dynamics.  (2005b, \S2)
\end{quote}
The arguments Greaves presents against this viewpoint seem to me to be convincing, but  for present purposes I shall simply presuppose that she is right. Wallace himself considers the possibility that SU might have to be rejected -- he calls the alternative \emph{Objective-determinism (OD)} -- and my objections focus on the version of the DW argument he takes to be relevant in that case.

\section{First concern}

A standard difficulty for symmetry-based probabilities is that different ways of carving up the relevant space of possible outcomes may yield different symmetries and hence different probabilities. In the classical case, there's scope to argue that one particular carve-up is to be preferred, precisely because it does yield assessments of probability that match the observed long(ish) run frequencies. But that response is off the menu, in the Everett case, for the very reason that's supposed to constitute the advantage, viz., that there is no unique outcome or observed frequency.

So the ``multiple carve-up'' problem would seem to be serious, for the DW argument, if other carve-ups were available. As Greaves (2004, \S5.2) notes, one other possibility might seem to be what she calls the ``Egalitarian'' proposal, which accords equal weight to all branches. In my view, Greaves is right about the problem this alternative would pose, if it were a genuine alternative. Fortunately for the DW argument, and as Greaves explains, it turns out not to be a serious possibility, for reasons to do with decoherence and the approximate nature of branches, amongst other things.\footnote{As Wallace puts it, `The point \ldots is not that there is no \emph{precise} way to define the number of
descendants \ldots. Rather, the point
is that there is \emph{not even an approximate way} to make such a
definition.'  (Wallace, 2005b, \S9; see also Greaves 2004, \S5.3)}

Still, the lesson seems to be that the DW argument would have trouble with the multiple carve-up problem, if Egalitarianism about branches were a live option -- if we shifted to a model in which branches were more sharply defined, as it were (and of some tractable cardinality, perhaps).\footnote{It is worth emphasising that what is excluded by this consideration is Egalitarianism \emph{about branches.} A different view would be Egalitarianism \emph{about outcomes} -- i.e., the view that for decision purposes we should accord equal weight  to all the \emph{outcomes} to which the Born rule ascribes non-zero probability. The lack of `even an approximate way' to define the number of descendants is bad news for  branch-Egalitarianism, but excellent news, at least \emph{prima facie,} for outcome-Egalitarianism -- for it suggests that among all the kinds of futures to which the Everett world commits us, there is no sense in which some occur ``more often'' than others. What better reason, surely, for treating them all on a par, for decision purposes? 
It is true that if we want to treat outcomes ``disjunctively'', dividing some finite measure between them, then it is difficult to say what counts as an outcome -- there are many alternative ways to partition the results of a measurement, in general. But as Peter Lewis has pointed out to me, this isn't a problem if we treat outcomes ``conjunctively'', taking seriously the idea that the implication of the Everett view is precisely that all possible outcomes are \emph{certain} to occur. In this case, we assign weight $1$ to all outcomes, and it doesn't matter how we partition them. In this sense at least, then, Egalitarianism does seem to remain in play -- an alternative decision policy whose exclusion needs to be justified, by a decision-based defence of the Born rule in the Everett context. }

If this is correct, it also seems to imply that the DW argument has similar trouble, if anyone can propose some reasonably elegant alternative to the Born rule as a rule weighting branches. If there's such alternative available, then it can't be true that it is uniquely rational to assign one's credences according to the Born rule -- after all, the same symmetry argument, starting with the alternative weights, would show that it is uniquely rational to use those weights, instead.

In her survey of these issues, Greaves (2004) suggests that we might turn this point around -- with  branch-
Egalitarianism off the menu, the Born rule might be the only non-trivial option available:

\begin{quote}
The failure of Egalitarianism (section 5.2) raises an important question: are 
there any coherent rationality strategies that are at all plausible and that also 
violate the Born rule? If not, we may hope to \ldots\ 
defend the Born rule by sheer process of elimination. (2004, \S5.4.2)
\end{quote} 

\noindent Greaves goes on to argue that no reasonable alternative is available: `It does seem that the only 
way we can come up with an [alternative] strategy is by brute force: that is, by specifying a preference ordering over Acts on a case-by-case basis, without 
appeal to any general governing rationale.'

On the face of it, however, there's an easy way to produce such alternatives: just use the weights provided by the Born rule itself, \emph{applied to a different initial state vector.} This could be done in a systematic way, apparently: we could imagine someone -- call her Heretic -- whose rule was that you started with the state vector $\ket{\psi}$, applied some given operator $\op{O}_{dd}$, and then used the Born rule on the resulting state vector $\ket{{\psi}'}=\op{O}_{dd}\ket{\psi}$. 

This may sound like a trivial suggestion. After all (it might be objected), the DW claim is only that \emph{if} you know the initial state vector, then your rational credences are uniquely constrained.\emph{ Of course} it follows that if we postulate a different initial state vector, we'll infer different rational credences.

But this misses the point of the objection. At the moment, what is at issue is the validity of the DW inference from a set of assumed weights to rational credences. We've seen that inference is challenged by the possibility of  `multiple carve-ups' -- different ways of assigning weights to branches or sets of branches. We repel one specific such challenge, by dismissing branch-
Egalitarianism, but the DW inference remains vulnerable to the general objection. And the present point is that the QM formalism itself seems to generate alternative weightings \emph{ad nauseum,} simply by considering alternative initial states.\footnote{We might want to impose a minimal consistency constraint, to the effect that the same branches receive non-zero weighting in the alternative state $\ket{{\psi}'}$ as in the original state $\ket{{\psi}}$.
Howard Barnum has pointed out to me that it isn't possible to meet this constraint in a perfectly general way. In an unpublished paper (Barnum, 1990) in which he considers a similar ``shuffling'' of the state vector, he shows that there is no non-trivial mapping $\op{O}_{dd}$ that my Heretic could use which preserves the extremal properties -- viz., that  the weight attached to a measurement result corresponding to an eigenvector orthogonal to the state must be zero, and that attached to getting a result corresponding to an 
eigenvector proportional to the state must be one. But Heretic could avoid this difficulty, presumably, by the simple expedient of announcing that her alternative recipe applies only in the cases in which the measurement in question is not extremal, in this sense, for either   $\ket{\psi}$ or $\ket{{\psi}'}$. As we'll see below, the question is really in what sense it could be rational to be guided by one quantum state rather than the other, in the Everett framework. To the extent that this is a serious issue,  it seems bad enough if it only ``bites'' in non-extremal cases. (Adam Elga has suggested to me an alternative way of dealing with the extremal cases, which he reports he  heard from Frank Arntzenius:
`don't identify ``zero amplitude'' with``does not occur''.  In other
words, think of things so that even when a term in the
preferred-basis-expansion of a state has amplitude zero, there \emph{is} a
branch associated with that term.  It's just a zero-weight branch -- cf.~the notion of a zero-mass particle.')
}

\subsection{The case against equivalence}

How do these considerations affect the DW argument? In Wallace's most recent development of the argument in (Wallace 2005b), a principle he  calls \textbf{equivalence} plays a crucial role. The objection outlined above appears to undermine the DW argument at precisely this point -- in other words, it undermines \textbf{equivalence}.

As Wallace explains (2005b, \S6), `\textbf{Equivalence} has the form of a principle of pure rationality: it dictates that any agent who does not regard equally-weighted events as equally likely is in some sense being irrational.'  He offers an argument for \textbf{equivalence}, which, as he stresses, makes essential use of the Everett interpretation:

\begin{quotation}
\noindent I wish to argue that the Everett interpretation necessarily plays 
a central role in any such defence [of \textbf{equivalence}]: in other interpretations, \textbf{equivalence} is
not only unmotivated as a rationality principle but is actually
absurd.\symbolfootnote[2]{[Wallace's footnote] ``In this section I confine my observations to those
interpretations of quantum mechanics which are in some sense ``realist''
and observer-independent (such as collapse theories or hidden-variable
theories). I will not consider interpretations (such as the Copenhagen
interpretation, or the recent variant defended by Fuchs and Peres 2000)
which take a more `operationalist' approach to the quantum formalism. It
is entirely possible, as has been argued recently by
Saunders (200[4]), that an approach based on Deutsch's proof may
be useful in these interpretations.''}

Why? Observe what \textbf{equivalence} actually claims: that if we know
that two events have the same weight, then we must regard them as
equally likely \emph{regardless of any other information we may have
about them}. Put another way, if we wish to determine which event to bet
on  and we are told that they have the same weight, we will be
uninterested in any other information about them.

But in any interpretation which does not involve branching --- that is,
in any non-Everettian interpretation --- there is a further
piece of information that cannot but be relevant to our choice: namely,
\emph{which event is actually going to happen}? If in fact we know that
$E$ rather than $F$ will actually occur, \emph{of course} we will bet on
$E$, regardless of the relative weights of the events. (Wallace 2005b, \S6)
\end{quotation}

\noindent Wallace's central argument for \textbf{equivalence} goes as follows:

\begin{quotation}
\noindent In fact, a very simple and direct justification of \textbf{equivalence}
is available. Consider, for simplicity, a Stern-Gerlach experiment \ldots : an atom is prepared in a
superposition $\ket{+_x}=\frac{1}{\sqrt{2}}(\ket{+_z}+\ket{-_z})$ and
then measured along the $z$ axis. According to the result of the
measurement, an agent receives some payoff. \emph{Ex hypothesi} the
agent is indifferent \emph{per se} to what goes on during the
measurement process and to what the actual outcome of the experiment is;
all he cares about is the payoff.

We now consider two possible games (that is, associations of payoffs
with outcomes):
\begin{description}
\item[Game 1:] The agent receives the payoff iff the result is spin up.
\item[Game 2:] The agent receives the payoff iff the result is spin
down.
\end{description}
In each game, the weight of the branch where the agent receives the
payoff is 0.5; \textbf{equivalence}, in this context, is then the claim that the agent is
indifferent between games 1 and 2.

To see that this is indeed the case, we need to model the games
explicitly. Let $\ket{\mbox{`up';reward}}$ and $\ket{\mbox{`down'; no reward}}$ be the quantum states of the two
branches on the assumption that game 1 was played: that is, let the
post-game global state\symbolfootnote[2]{[Wallace's footnote] ``More precisely: the global state relative to the pre-game agent: there
are of course all manner of other branches which are already effectively disconnected from the agent's branch.''} if game 1 is played be
\be
\ket{\psi_1}=\frac{1}{\sqrt{2}}(\ket{\mbox{`up';reward}}+ \\ \ket{\mbox{`down'; no reward}}).\ee
Similarly, if game 2 is played then the quantum state is
\be
\ket{\psi_1}=\frac{1}{\sqrt{2}}(\ket{\mbox{`down';reward}}+ \\ \ket{\mbox{`up'; no reward}}).\ee
Why should an agent be indifferent between a physical process which
produces \ket{\psi_1} and one which produces \ket{\psi_2}?

Well, recall that the agent is indifferent \emph{per se} to the result
of the experiment. This being the case, he will not object if we erase
that result. Let \ket{\mbox{`erased',reward}} indicate the state of the
branch in which the reward was given  post-erasure and \ket{\mbox{`erased',no reward}}
the post-erasure state of the no-reward branch. Then (game 1+erasure)
leads to the state
\be
\ket{\psi_{1;e}}=\frac{1}{\sqrt{2}}(\ket{\mbox{`erased',reward}}+\ket{\mbox{`erased',no
reward}})\ee
and (game 2+erasure) to the state
\be
\ket{\psi_{2;e}}=\frac{1}{\sqrt{2}}(\ket{\mbox{`erased',reward}}+\ket{\mbox{`erased',no
reward}})\ee
--- that is, (game 1+erasure) and (game 2+erasure) lead to the same
state. If (game 1+erasure) and (game 2+erasure) are just different ways
of producing the same physical state --- different ways, moreover, which
can be made to differ only over a period of a fraction of a second, in
which the agent has no interest --- then the agent should be indifferent
between the two. Since he is also indifferent to erasure, he is
indifferent between games 1 and 2, as required by \textbf{equivalence}. (Wallace 2005b, \S9)
\end{quotation}

Against this background, consider our Heretic. Suppose for the sake of argument that while agreeing that the initial state of the atom is the 
superposition $\ket{\psi}=\ket{+_x}=\frac{1}{\sqrt{2}}(\ket{+_z}+\ket{-_z})$, she weights the branches for decision theory purposes as if it were  $\ket{\psi{'}}=\frac{1}{\sqrt{3}}\ket{+_z}+\frac{\sqrt{2}}{\sqrt{3}}\ket{-_z}$. (In other words, she assigns subjective probabilities in the way that an Orthodox person would, following Wallace's principle \textbf{equivalence}, if the initial state were actually  $\ket{\psi{'}}$.)

Clearly, Heretic will prefer Game 2 to Game 1, and will be (rightly) unmoved by the consideration that once the chosen game is over, thanks to erasure, she will be unable to tell whether it was actually Game 1 or Game 2 that was played. In one sense, of course, she is equally happy to win by either path. But she thinks she is twice as likely to win by Game 2 as by Game 1, so she should not weight them equally, for decision purposes, \emph{contra} \textbf{equivalence}.

This means that unless it has \emph{already been assumed,} implicitly, that rational choice is constrained by the Born rule, then it is perfectly possible to make sense of someone who violates \textbf{equivalence}. At least, it is possible to do so \emph{modulo} the assumption that rational choice of this kind makes sense at all, in the Everett context, but DW can hardly dispute that assumption, in this context. 

To put it another way, the objection seems to establish the following: if we assume that rational choice under uncertainty makes sense in the Everett context, and don't assume what we are trying to establish -- viz., that rational choice should be guided by probabilities derived from the Born rule -- then it is easy to make sense of a rational agent whose (symmetry-driven) credences differ from those assigned by the Born rule, and violate \textbf{equivalence}.

So the multiple carve-up problem seems  to be a real difficulty. It is no help knowing the real initial state vector, because the symmetry argument that is supposed to justify the Born rule is already undermined by the existence of alternative carve-ups. What would be needed would be some primitive assumption to the effect that carve-up given by the real state vector was privileged. However,  this seems to amount to assuming by fiat what the DW argument was supposed to show.

In fact, the consequences may be even worse than this. If the DW argument worked, it would seem to establish not only the Born rule, but also, more basically, the coherence of rational decision under uncertainty in the Everett context. If the objection above is correct in claiming that the Born rule is effectively assumed as a primitive in the DW argument, then the effect isn't simply to undermine the claim to have \emph{derived} the Born rule; it is to challenge the coherence of rational decision in an Everett world, in a fairly precise sense. What it challenges is the idea that there could be any rational basis for choosing between the weights associated with alternative initial state vectors, as one's guide to such decisions.\footnote{At least within the scope of the variation permitted by the considerations noted in fn.\ 3 above. } 

The upshot seems to be something like this: in the Everett context we can make sense of a minimal coherence condition of the kind provided by Dutch book argument -- failing which, apparently, an agent can be made to be worse off in all future branches.\footnote{Though it is worth asking to what extent these arguments themselves depend on the assumption that well-informed agents can make sense of choice of betting quotients, in the Everett context.}  But there seems to be no  basis for any rational preference between credences beyond that -- no reason for choosing one initial state vector rather than another, as it were, if one's weightings are to be given by the Born rule. 

If this is correct, then even the proposal to assume the Born rule as a primitive (as recommended by Greaves 2004, for example) seems to be in trouble. We have not been told what the content of such an assumption would be -- what it would be for the assumption to be correct, or what difference it makes. It is no use saying that its content consists precisely in the way that it guides our choices in cases of decision under uncertainty. The point is that we haven't been given any understanding of what difference those choices make (within the circle of the alternatives permitted by coherence in the minimal sense); or of what it could amount to for the assignment according to the Born rule to be the ``right'' assignment.

Another way to put this is to say that if the link between the quantum amplitudes and decision probabilities is taken as primitive, then we lack an sense as to how there could be a single, correct initial state vector, in a given situation. It doesn't seem to be helpful to say that this is simply a postulate of the theory. When we have real frequencies of outcomes, we have some sense of what correctness might mean. Here, the point is that this sense seems to have gone missing. All possibilities seem to be on a par (extremal cases aside, perhaps).

It is natural to think that if we are to find a solution to this problem, we will need to look to the epistemology of the Everett model -- at this point, surely, we should hope to find something that will distinguish alternative hypotheses about the initial state. My second concern, below, focusses on this issue. Before coming to that, however, I want to raise another issue about Wallace's argument for \textbf{equivalence}.

\subsection{Same-State Rational Indifference}

As we have seen, Wallace's argument for \textbf{equivalence} relies on an argument which first appears in this passage:

\begin{quotation}\noindent
[R]ecall that the agent is indifferent \emph{per se} to the result
of the experiment. This being the case, he will not object if we erase
that result. Let \ket{\mbox{`erased',reward}} indicate the state of the
branch in which the reward was given  post-erasure and \ket{\mbox{`erased',no reward}}
the post-erasure state of the no-reward branch. Then (game 1+erasure)
leads to the state
\be
\ket{\psi_{1;e}}=\frac{1}{\sqrt{2}}(\ket{\mbox{`erased',reward}}+\ket{\mbox{`erased',no
reward}})\ee
and (game 2+erasure) to the state
\be
\ket{\psi_{2;e}}=\frac{1}{\sqrt{2}}(\ket{\mbox{`erased',reward}}+\ket{\mbox{`erased',no
reward}})\ee
--- that is, (game 1+erasure) and (game 2+erasure) lead to the same
state. If (game 1+erasure) and (game 2+erasure) are just different ways
of producing the same physical state --- different ways, moreover, which
can be made to differ only over a period of a fraction of a second, in
which the agent has no interest --- then the agent should be indifferent
between the two. Since he is also indifferent to erasure, he is
indifferent between games 1 and 2, as required by \textbf{equivalence}. (Wallace 2005b, \S9)
\end{quotation}
Let's focus on the principle appealed to in this passage, which we might formalise as follows:

\begin{description}
\item[\textbf{Same-State Rational Indifference (SSRI):}] A rational agent should be indifferent
between two options that give rise to the same global state (relative to the pre-game agent).
\end{description}

In the previous subsection, we saw how our agent Heretic seemed to have grounds for rejecting SSRI: believing she had a better chance of winning via Game 2 than Game 1, she prefers Game 2, despite the fact that the two games lead to (effectively) the same global final state. The point I want to make now is that Heretic's case illustrates a general reason why SSRI must be rejected, by anyone who wants to make sense of (some analogue of) rational decision under uncertainty, in an Everett world. If a rational agent's pre-game preferences are constrained by the global final state in this way, she is bound to be indifferent between the possible (local, in-branch) \emph{outcomes} of the game or measurement in question, because all of them reflect the same global final state. 

SSRI thus ensures that  a rational Everettian agent is indifferent between the possible outcomes of any bet whatsoever (once the die is cast, at any rate)!  This may be a prescription for a certain (Buddhist) sort of contentment, but it seems to be a \emph{reductio} of the attempt to apply the decision calculus to the Everett world. Without differential preferences between outcomes, betting behaviour cannot track credences. As it stands, then, SSRI seems to be a ``home-goal'' for the project of making sense of rational decision in the Everett context. 

Does Wallace's argument for  \textbf{equivalence} actually require SSRI? The obvious thought is that we might avoid it by arguing for the required indifference between Game 1 and Game 2 not in terms of the equivalence of the global states resulting from each game, but in terms of the  pairwise equivalence of the two possible outcomes of each game. Thus, modifying Wallace's terminology in the obvious way, we would argue that the agent should be indifferent between the post-erasure winning outcomes of Game 1 and Game 2, viz., $\ket{\mbox{`erased',reward}}_1$ and $\ket{\mbox{`erased',reward}}_2$, respectively; and also between the corresponding losing outcomes, $\ket{\mbox{`erased',no reward}}_1$ and $\ket{\mbox{`erased',no reward}}_2$, respectively. 

At this point, we want to conclude that the agent should hence be indifferent between Game 1 and Game 2. However, it seems obvious that this depends on the very matter at issue: namely, whether the agent should assign equal weight, \emph{for such decision purposes,} to the two different measurement outcomes on which success depends, in the two versions of the game. 

In essence, then, it seems to me that Wallace's argument seeks to avoid circularity by moving to the ``global'' level; but that at that level it invokes a principle (SSRI) which  defeats the general project of making sense of rational decision under uncertainty in the Everett world.\footnote{This doesn't imply that the SSRI isn't defensible in the Everett context, of course.}

\section{Second concern}

As I noted above, it is natural to think that if we are to find a solution of what I termed the multiple carve-up  problem, we will need to look to the epistemology of the Everett model. It is at this point that we might hope to find something to distinguish alternative hypotheses about the initial state. But this brings me to my second concern. 

Suppose I accept the Everett model. Whatever outcome of a quantum measurement I observe in ``my'' branch, I should believe that each of the other possible outcomes has occurred in some other branch. In other words, I should take it that what I have observed is only one part or aspect of the overall outcome of the process in question. Doesn't the principle of total evidence require me to take that into account, at this point? If so, then -- in the absence of any further facts about \emph{relative frequencies} of the various outcomes -- there seems to be no way in which the evidence can confirm any particular hypothesis about the relative weights of the various branches. For the total evidence (in this sense) is exactly the same in all cases. (In a sense, this appeals to a kind of epistemological analogue of SSRI.)

It might be objected that this argument ignores the indexical component in ``my'' evidence, and especially my normal entitlement to regard myself as typical -- this entitlement, the objector would say, enables ``me'' to infer something further about the total evidence, viz., that it is likely to be similar to the evidence ``I'' actually have. But the notion of typicality rests on frequency comparisons. If such comparisons don't make sense in the Everett world, as Wallace and Greaves maintain, this response seems to be off the agenda.

\subsection{Greaves on Everettian epistemology}

Greaves herself offers a defence of the epistemology of the Everett model. Her argument is based on the distinctive interpretation she proposes of the Born probabilities, as the weightings that a rational agent should give to the interests of each of her multiple descendants in an Everett world -- as she puts it, the Born probabilities provide a `caring measure'. 

On the face of it, this proposal encounters the difficulty identified above: if the practical significance of the state vector is to be linked to a caring measure in this way, we are entitled to ask what it means for one particular state to be the ``correct'' state -- how does the ontology of the Everett picture make caring measure better than another. (Again, a single-history ontology has some prospect of an answer in terms of actual frequencies, but that option isn't available here.)

Greaves's proposal, in effect, is that we can get the constraint we need from epistemology. She argues that the ``caring'' framework provides a close analogue of conditionalization, sufficient to ensure we have good reason to afford high credence to the conventional quantum state ascriptions,   given the abundance of experimental evidence already accumulated. In her own words, she proposes that 
`a fairly strong case can be made for the following philosophical conjecture:
\begin{quote}
If the rational Everettian cares about her future successors in proportion to their relative amplitude-squared measures, then (given our 
actual experimental data) she should regard (Everettian) quantum 
mechanics as empirically confirmed.' (2004, \S2.3)
\end{quote}
If this argument succeeds, it  represents progress. It provides an \emph{epistemological} sense in which a particular initial quantum state might reasonably be privileged -- regarded as distinctively ``correct'' -- by observers in our situation. With that sense in place, it might be argued that the Orthodox proposal to base one's decision probabilities (or caring measure) on this preferred state (via the Born rule) is at least \emph{simpler} than Heretic's proposal to begin with a different initial state (systematically derived from the ``correct'' initial state).

However, there seem to be two difficulties with Greaves's proposal. 

\subsubsection{First difficulty: Naive Conditionalization}

The first difficulty turns on the objection outlined at the beginning of this section. Consider an agent who accepts the Everett view, and is interested in testing or confirming an hypothesis about the initial state of a quantum coin-tosser. He sets the device working, and sits down to record the results. 
He observes Heads, reasons that Tails has occurred in the other branch, and concludes that the total state of the world is exactly what he predicted it would be, with certainty, before the experiment. Since this reasoning is insensitive to the weight his hypothesis ascribes to Heads, his observations cannot bear on that matter, no matter how much data he collects. But if the suggestion above about the role of the principle of total evidence is correct, this is precisely the kind of agent we ought to be. If so, then from an Everettian point of view, our accumulated experimental data tells us very little about the ``true'' quantum state.

Greaves considers a related point in a recent paper (Greaves 2006). Again, her project  is to argue that an orthodox epistemic updating rule works the same in the Everett case as in conventional ``one-world'' QM (so that existing experimental evidence provides the same degree of support for the orthodox quantum state ascriptions). She considers an alternative updating policy -- `Naive Conditionalization', as she calls it -- closely related to that of the agent we have just considered.\footnote{The main difference is that Greaves is considering the case in which  the Everett view itself is treated as an hypothesis under consideration, rather than, as above, as a presupposition of an hypothesis about the correct quantum weights in a particular measurement system.}

Naive Conditionalization rests on the thought that since the Everett view implies that any measurement outcome with non-zero probability is bound to occur in some branch,  
\begin{quote}
\emph{every} event that is quantum-mechanically possible, but not necessary, confirms QME [i.e., the Everett view] at the expense of theories that assign probability less than unity to that event. (2006, \S 4)
\end{quote}
As Greaves points out, 
\begin{quote}
[t]his result is not really surprising: what we have done, in constructing Naive Conditionalization, is to give formal expression to the intuition that disconfirmation of a theory occurs only when something that, according to the theory in question, is \emph{improbable,} occurs; and (at the same time) that, according to QME, nothing is improbable since everything is certain to occur.
\end{quote}
She concludes:
\begin{quote}
This observation is (presumably!) a \emph{reductio} of the suggestion that Naive Conditionalization is the rational updating policy when branching-universe theories are among those under consideration. 
\end{quote}

These are difficult issues, but it seems to me that Naive Conditionalization has more to be said for it than Greaves allows. By way of analogy, consider this variant of the Sleeping Beauty problem.\footnote{For the standard version of the problem, see Elga 2000, Lewis 2001,  Vaidman and Saunders 2001, Dorr 2002, Arntzenius 2003, Horgan 2004 Weintraub 2004 and
White 2006, for example. I have modified the standard problem by removing a temporal asymmetry which is distracting for present purposes, and by making the numbers bigger. (Thanks to Jenann Ismael for calling my attention to the connection between Sleeping Beauty and the present issue.)} An experimental subject, Beauty, is informed that on the basis of a toss of a fair coin, she is to be assigned to one of two test groups, H(eads) and T(ails),  but not told which. She is told that subjects in Group H will be put to sleep from Sunday evening to the following Saturday morning, and will be woken once, on one of the five intervening days; the particular day being chosen at random for each subject individually. Subjects in Group T will also be put to sleep from Sunday evening to the following Saturday morning, but will be woken on each intervening day; and given a drug to ensure that they cannot remember previous wakings, if any. Beauty wakes one morning, and discovers that it is Thursday. What credence should she give to  the proposition that she is in Group T? 

There is strong intuitive appeal to the thought that since \emph{all} of Group T but only 20\% of Group H find themselves in Beauty's situation, the odds are strongly in favour of her being in Group T. Yet the inference seems formally parallel to Greaves's Naive Conditionalization. Someone who wishes to reject  Naive Conditionalization needs either (i) to reject the corresponding inference in the Sleeping Beauty case (arguing that Beauty should give credence $0.5$ to  the proposition that she is in Group T); or (ii) to find some relevant epistemic difference between the two cases.

To make the parallels more explicit, imagine a quantum-theoretic version of the Sleeping Beauty experiment (the Sleeping QT experiment, perhaps). In this case, both groups of subjects  -- H$_Q$ and T$_Q$, let's say  -- are woken only once. In  Group T$_Q$ only, however, the waking day is chosen for each subject individually according to the result of a  quantum measurement with five equally weighted outcomes. (In Group H$_Q$, perhaps, it is chosen by consulting some list of effectively random numbers, fixed in advance, in such a way that all five possibilities are equally likely, at least approximately.) In addition, subjects are informed (or led to believe) that the Everett view is correct, and made to understand the consequences, in the present case: in effect, each of the subjects in Group  T$_Q$ will branch into five descendants (or subsets of descendants), one for each possible choice of waking day; whereas those in  Group H$_Q$ will not branch in this way. 

A particular subject, Na\"{i}f, wakes to find that it is Thursday. He follows Beauty's lead, reasoning  that since it is five times less likely that the world contains a Na\"{i}f-waking-on-Thursday event if he is in Group  H$_Q$ than if he is in Group T$_Q$, he is probably in the latter. What, if anything, has he got wrong?\footnote{For the sake of this issue, obviously, it doesn't matter whether the Everett view is actually correct. We are interested in how one should reason if one believes that it is correct.}

It might be suggested that we should consider the case in which Na\"{i}f doesn't know immediately what day it is, when he wakes up. Then in either case -- whether he is in Group H$_Q$ or Group T$_Q$ -- he seems correct to assign credence $0.2$ to the possibility that it is Thursday.\footnote{In the Group T$_Q$ case, of course, this depends on whether he is right to assign equal credence to options with equal quantum amplitudes, but let's let that pass, for now. For the moment, the question is whether an Everettian agent who accepts the Born rule should update according to probabilities it supplies -- the Naive Conditionalizer says not.} Doesn't this  suggest that at this point, he has no reason to favour either hypothesis about which group he is in?   And if that's right, then how could the discovery that it actually is Thursday make such a difference? (After all, he knew in advance what the  possibilities were.) 

But consider Beauty's case. She, too, can be in this position of ignorance, if she is not told what day it is, when she first wakes up. She, too, should assign equal weight to all five possibilities. But this doesn't mean that she has no reason to favour the hypothesis that she is in Group T. On the contrary,  apparently, she knows that whatever the answer is, it favours the hypothesis that she is in Group T -- whatever day it is, her situation is five times more likely to have arisen in Group T than in Group H. 
And the same goes for Na\"{i}f, or so it seems.\footnote{Except that in his case, as just noted, we still lack a reason for assigning particular values to the ignorance-based probabilities.} He, too, can reason that whatever outcome he discovers when he inspects the result of the experiment (i.e., when he discovers what day it is), the existence of someone ``in his shoes'' discovering that outcome will be five time more likely in Group T$_Q$ than in Group H$_Q$. 

The right diagnosis seems to be the following. It is true that at this point in Na\"{i}f's deliberations -- i.e., after he wakes up, but before he is told that it is Thursday -- P(Thursday$|$T$_Q$) = P(Thursday$|$H$_Q$), and hence that updating on the new information that it is Thursday will make no further difference to his rational credence in T$_Q$ and H$_Q$. But by this point, his rational credences have \emph{already} been shifted in favour of T$_Q$, by the discovery that on this day, whatever it is, there is someone ``in his shoes'' who is awake. That state of affairs is much more likely  in Group T$_Q$ than in Group H$_Q$. (Once again, the same diagnosis seems to apply in Beauty's case, \emph{mutatis mutandis.)}

As I said, these are difficult matters, and they  deserve a more detailed treatment than I can give them here.\footnote{I'm grateful to Peter Lewis for showing me recent work in which he, too, has been exploring the implications of Sleeping Beauty for Everett probabilities, and  reaching similar conclusions; see Lewis (2006).} Provisionally, however, it seems to me that a case has been made for the claim that  Naive Conditionalization \emph{i}s the correct policy in the Everett world (or when the Everett view is one of the options). If so, then Greaves may be right to conclude that this `is (presumably!) a \emph{reductio}', but wrong about the target of the \emph{reductio.} The problem isn't with Naive Conditionalization -- which is just ordinary conditionalization, applied in the case in which the relevant conditional probabilities go to unity -- but with the Everett view itself. (In fact, it  seems to precisely the kind of problem we would expect anyway, if we were offered any  more trivial theory which claimed to imply that ``all possibilities actually happen'' -- it doesn't seem to be an advantage of such a theory that anything confirms it!) 

Perhaps it is too strong to describe this as a \emph{reductio} of the Everett proposal.\footnote{Even if one agrees that Naive Conditionalization is the right policy, given the Everett view.} It may be more accurate to say that the upshot is only that the Everett view belongs to a class of theories that are inevitably ``pathological'' with respect to standard Bayesian confirmation. There's then room for argument about the significance of this fact.\footnote{Compare  the conclusion that a particular theory is unfalsifiable, in a Popperian sense. Few people would regard this as an epistemological virtue, but it isn't usually thought to be a \emph{reductio,} as such.} However, in case anyone is inclined to regard it as a mark in favour of the Everett view that it gets ``confirmed'' so easily, it is worth pointing out that the effect is independent of the Born probabilities. Hence it is no help at all with the problem raised in the previous section, of giving physical meaning to the state ascription.

\subsubsection{Second difficulty: epistemology depends on decision}

The second difficulty with Greaves's proposal emerges in the light of the structure of the argument in this note. In effect, our first concern, in \S2, offered a sceptical conclusion about decision-theoretic reasoning in an Everett world. We argued that from a decision-theoretic point of view, the Everett model provides no sense in which -- if we do allow our decision probabilities (or caring measure) to be guided by the Born rule -- we could think that one of a range of initial states was any better than another, as the input to the Born rule. A sceptic in the grip of this conclusion lacks any sense of what it means for a particular choice of betting quotients to be ``right'', or ``appropriate'', over and above the minimal consistency constraints required for coherence.\footnote{Note that such a sceptic is not the same character as our Heretic from \S2, who follows an alternative to the Born rule. Heretic thinks that the Born rule is \emph{wrong,} and proposes an alternative. Sceptic -- moved, perhaps, by consideration of Heretic's case -- claims to lack any sense of what it means for the Born rule to \emph{be} ``right'' or ``wrong''.} 

The hope was to alleviate the sceptic's decision-theoretic nihilism by appealing to the epistemology of the Everett interpretation. By Greaves's path, however, the relevant epistemology depends squarely on the decision theory. The defence of conditionalization depends on a Dutch Book argument, just as it does in the classical case; but the Dutch Book argument cannot get off the ground, unless the agent concerned can be assumed to have betting quotients. It is not clear that that's true, of our Sceptic. 

Thus the structure of the second difficulty for Greaves's proposal is as follows: in the present context, we need to make sense of the epistemology of the Everett view, to provide a necessary foundation for its decision theory. But Greaves's route to the epistemology depends on a decision-theoretic argument. Hence it seems inadmissable, as a solution to the sceptical difficulty identified in \S2.

\section{Conclusion}

The two concerns above are related, apparently, because both stem from the feature that Wallace identifies as the source of the unique power of the appeal to symmetry in the Everett case, viz., that there is no unique outcome to spoil the symmetry. In both cases, the objection is that in the absence of frequencies, that feature inevitably severs the link between weights and outcomes, and puts all of a wide class of assignments of weights on a par (predictively, for the purposes of action, in the first case; and retrodictively, for the purpose of epistemology and confirmation, in the second).

To give the penultimate word to Greaves herself, 
\begin{quote}
the worry is this: it may be that if \ldots\  we decide to understand quantum mechanics along Everettian lines,  \ldots\  we lose the empirical reason we had for believing \emph{quantum mechanics} in the first place. (2004, \S2.2)
\end{quote} 
I've argued that this worry is more serious than Wallace and Greaves realise. In the version of Everett  to which they both subscribe, in which there are no helpful facts about relative frequencies of branches, the internal symmetries of quantum mechanics generate a fatal indeterminacy. The central ontological fact of the Everett world, viz., the identity of the quantum state itself, seems to lose practical and empirical significance, within wide margins.\footnote{I am much indebted  to  Guido Bacciagaluppi, Howard Barnum, Adam Elga, Jenann Ismael,  Chris Fuchs and Peter Lewis, and especially to Hilary Greaves herself, for many helpful comments on previous versions of this note. I am also grateful to the Australian Research Council and the University of Sydney, for research support.}

\section*{References}

  Arntzenius. F. (2003). Some problems for conditionalization and reflection. \emph{Journal of Philosophy 100,} 356--370.

\vspace{6pt}\noindent  Barnum, H. (1990). The many-worlds interpretation of quantum mechanics: psychological versus physical bases for the multiplicity of ``worlds''. Available online at \href{http://philsci-archive.pitt.edu/archive/00002647/}{http://philsci-archive.pitt.edu/archive/00002647/}.

\vspace{6pt}\noindent  Barnum, H., C.~M. Caves, J.~Finkelstein, C.~A. Fuchs, and R.~Schack (2000).
 {Q}uantum {P}robability from {D}ecision {T}heory?
 {\em Proceedings of the Royal Society of London\/}~{\em A456},
  1175--1182.
 Available online at \\\href{http://www.arXiv.org/abs/quant-ph/9907024}{http://www.arXiv.org/abs/quant-ph/9907024}.
 
\vspace{6pt}\noindent  Deutsch, D. (1999).
 {Q}uantum {T}heory of {P}robability and {D}ecisions.
 {\em Proceedings of the Royal Society of London\/}~{\em A455},
  3129--3137.
 Available online at \href{http://www.arxiv.org/abs/quant-ph/9906015}{http://www.arxiv.org/abs/quant-ph/9906015}.

\vspace{6pt}\noindent  Dorr, C. (2002). Sleeping Beauty: in defence of Elga. \emph{Analysis 62,} 292--296.
 
\vspace{6pt}\noindent  Elga, A. (2000). Self-locating belief and the Sleeping Beauty problem. \emph{Analysis 60, }143--47.
 
 \vspace{6pt}\noindent  Fuchs, C. and A.~Peres (2000).
 Quantum theory needs no ``interpretation''.
 {\em Physics Today\/}~{\em 53\/}(3), 70--71.

\vspace{6pt}\noindent Greaves, H. (2004).
 Understanding {D}eutsch's probability in a deterministic multiverse.
 {\em Studies in the History and Philosophy of Modern Physics\/}~{\em
  35}, 423--456.
 Available online at \href{http://xxx.arXiv.org/abs/quant-ph/0312136}{http://xxx.arXiv.org/abs/quant-ph/0312136}.

\vspace{6pt}\noindent Greaves, H. (2006).
 Probability in the Everett interpretation: a solution to the epistemic problem. Preprint, available online at\\ \href{http://www.rci.rutgers.edu/~hgreaves/papers/epistemic.pdf}{http://www.rci.rutgers.edu/$\sim$hgreaves/papers/epistemic.pdf}.

\vspace{6pt}\noindent Horgan, T. (2004). Sleeping Beauty awakened: new odds at the dawn of the new day. \emph{Analysis 64,} 10--21.

\vspace{6pt}\noindent Lewis, D. (2001). Sleeping Beauty: reply to Elga. \emph{Analysis 61,} 171--176.
  
\vspace{6pt}\noindent   Lewis, P. (2003).
 {D}eutsch on quantum decision theory.
 Available online from \href{http://philsci-archive.pitt.edu/archive/00001350/}{http://philsci-archive.pitt.edu/archive/00001350/}.
 
\vspace{6pt}\noindent   Lewis, P. (2005).
 Probability in Everettian quantum mechanics.
 Available online from \href{http://philsci-archive.pitt.edu/archive/00002716/}{http://philsci-archive.pitt.edu/archive/00002716/}.

 \vspace{6pt}\noindent   Lewis, P. (2006).
Quantum Sleeping Beauty. Available online from \\\href{http://philsci-archive.pitt.edu/archive/00002715/}{http://philsci-archive.pitt.edu/archive/00002715/}.

 \vspace{6pt}\noindent  Saunders, S. (2004).
 Derivation of the {B}orn rule from operational assumptions.
 \emph{Proceedings of the Royal Society A 460,} 1--18.
  Available online from \href{http://xxx.arxiv.org/abs/quant-ph/0211138}{http://xxx.arxiv.org/abs/quant-ph/0211138}.
  
   \vspace{6pt}\noindent  Vaidman, L. and Saunders, S. (2001). On Sleeping Beauty Controversy. Available online at \href{http://philsci-archive.pitt.edu/archive/00000324/}{http://philsci-archive.pitt.edu/archive/00000324/}.
  
 \vspace{6pt}\noindent  Wallace, D. (2003).
{E}verettian rationality: defending {D}eutsch's approach to
  probability in the {E}verett interpretation.
{\em Studies in the History and Philosophy of Modern Physics\/}~{\em
  34}, 415--439.
Available online at \\\href{http://xxx.arXiv.org/abs/quant-ph/0303050}{http://xxx.arXiv.org/abs/quant-ph/0303050}.
  
\vspace{6pt}\noindent Wallace, D. (2005a). Epistemology quantized: circumstances in which we should come to  believe in the Everett interpretation.  Available online at\\
  \href{http://philsci-archive.pitt.edu/archive/00002368/}{http://philsci-archive.pitt.edu/archive/00002368/}.
  
\vspace{6pt}\noindent Wallace, D. (2005b).   Quantum Probability from Subjective Likelihood: improving on
Deutsch's proof of the probability rule. Available online at\\
  \href{http://philsci-archive.pitt.edu/archive/00002302/}{http://philsci-archive.pitt.edu/archive/00002302/}.
  
\vspace{6pt}\noindent Weintraub, R. (2004). Sleeping Beauty: a simple solution. \emph{Analysis 64,} 8--10.

\vspace{6pt}\noindent White, R. (2006). The generalized Sleeping Beauty problem: a challenge for thirders. \emph{Analysis 66,} 114--119.

 \end{document}